\documentclass[prl,aps,amsmath,amssymb, preprintnumbers, reprint,longbibliography,superscriptaddress,nofootinbib]{revtex4-2}
\usepackage{amsmath}
\usepackage{amssymb}    % 
\usepackage{amsthm}    % 
\usepackage{bm}    % 
\usepackage{epsfig}
\usepackage{hyperref}
\usepackage{subfigure}
\usepackage{float}
\usepackage{verbatim} 
\usepackage{pstricks}
\usepackage{color}
\usepackage{slashed}
\usepackage{multirow}
\usepackage{pstricks}
\usepackage{booktabs}
\usepackage{subfigure}
\usepackage{epstopdf}
\usepackage{braket}
\usepackage{blkarray}
\usepackage{enumerate}
\usepackage{footnote}

\allowdisplaybreaks[3]

\newcommand{\bea}{\begin{eqnarray}}
\newcommand{\eea}{\end{eqnarray}}
\usepackage[normalem]{ulem} 
\newcommand{\mc}{\mathcal}

\usepackage{ulem}

\begin{document}
\preprint{
	{\vbox {
			\hbox{\bf MSUHEP-24-006}
}}}
\vspace*{0.2cm}

\title{Lam-Tung relation breaking in $Z$ boson production as a probe of SMEFT effects}
\author{Xu Li}
\email{lixu96@ihep.ac.cn}
\affiliation{Institute of High Energy Physics, Chinese Academy of Sciences, Beijing 100049, China}
\affiliation{School of Physical Sciences, University of Chinese Academy of Sciences, Beijing 100049, China
}
\author{Bin Yan}
\email{yanbin@ihep.ac.cn (corresponding author)}
\affiliation{Institute of High Energy Physics, Chinese Academy of Sciences, Beijing 100049, China}

\author{C.-P. Yuan}
\email{yuan@pa.msu.edu}
\affiliation{
	Department of Physics and Astronomy, Michigan State University, East Lansing, MI 48824, USA
}

\begin{abstract} 
The violation of Lam-Tung relation in the high-$p_T^{\ell\ell}$ region of the Drell-Yan process at the LHC presents a long-standing discrepancy with the standard model prediction at $\mathcal{O}(\alpha_s^3)$ accuracy. In this Letter, we employed a model-independent analysis to investigate this phenomenon within the framework of the Standard Model Effective Field Theory (SMEFT). Our findings revealed that the leading contributions from SMEFT to this violation appear at the $1/\Lambda^4$ order with $\mathcal{O}(\alpha_s)$ accuracy in QCD interaction. Notably, we demonstrated that the quadratic effect of  dimension-6 dipole operators, associated with the $Z$ boson, dominates the breaking effects induced by various dimension-6 and  dimension-8 operators. This provides a possible explanation for the observed discrepancy with the Standard Model predictions at the LHC. Furthermore, the breaking effects could also serve as a powerful tool for constraining $Z$-boson dipole interactions, highlighting their importance among potential sources of new physics in the Drell-Yan process.
\end{abstract} 
\maketitle

%================================================================================================
\noindent {\bf Introduction.}
%================================================================================================
The dilepton angular distributions in the Drell-Yan process at the Large Hadron Collider (LHC) directly probe the polarization effects of the gauge bosons $Z/\gamma^\star$ and can provide key information of the electroweak interactions between gauge bosons and fermions in the Standard Model (SM) and beyond. The well-known Lam-Tung relation for the angular coefficients $A_0=A_2$~\cite{Lam:1978pu,Lam:1978zr,Lam:1980uc}, derived from the angular distributions of lepton pair, is a crucial observable in the analysis of the subtle QCD and electroweak effects. This relation can hold up to $\mathcal{O}(\alpha_s)$ in perturbative QCD under the leading twist approximation, and is a consequence of the spin-1/2 nature of the quarks at the tree-level, and vector coupling feature of spin-1 gluon to quarks at $\mathcal{O}(\alpha_s)$ accuracy~\cite{Lam:1978pu,Lam:1978zr,Lam:1980uc}. It has been demonstrated that the breaking
of the Lam-Tung relation can be emerged due to the non-coplanarity between the hadron plane and parton plane~\cite{Peng:2015spa}, which can occur at the $O(\alpha_s^2)$ and beyond in perturbative QCD. The breaking effects of this relation were confirmed by both the ATLAS~\cite{ATLAS:2016rnf}  and CMS~\cite{CMS:2015cyj} collaborations at a centre-of-mass energy of $\sqrt{s}=8$ TeV and LHCb~\cite{LHCb:2022tbc} collaboration at $\sqrt{s}=13$ TeV. However, the experimental values of the regularized data show a significant deviation from the prediction of the SM at $O(\alpha_s^2)$ in the high-$p_T^{\ell\ell}$ region~\cite{ATLAS:2016rnf}, where $p_T^{\ell\ell}$ denotes the transverse momentum of the lepton pair, and a systematic deviation is still observed compared to the $O(\alpha_s^3)$ predictions for $p_T^{\ell\ell}>50$ GeV~\cite{Gauld:2017tww}. Despite the inclusion of additional electroweak corrections, their effects still cannot resolve this long-standing issue in Drell-Yan process~\cite{Frederix:2020nyw}. 
However, nonperturbative effects, such as the contributions from higher twist~\cite{Balitsky:2021fer} and parton transverse momentum effects~\cite{Brzeminski:2016lwh,Motyka:2016lta,Nefedov:2020ugj}, could be a potential source of the observed discrepancy in the Lam-Tung relation. Therefore, the study of this observable in the Drell-Yan process is widely regarded as a crucial avenue for delving into the fundamental properties of QCD.

In this Letter, we present the first investigation into the possibility that the discrepancy in the Lam-Tung relation from the high-$p_T^{\ell\ell}$ region at the LHC is attributed to new physics (NP) beyond the SM. With no hint of any new heavy particles at the LHC, the best strategy to search for NP effects is to use the standard model effective theory (SMEFT) to systematically parametrize the ignorance of UV physics~\cite{Buchmuller:1985jz,Grzadkowski:2010es}. Our investigation reveals that the breaking of the Lam-Tung relation first occurs at the $O(\alpha_S/\Lambda^4)$, with $\Lambda$ representing the scale of the NP. This implies that the linear effects of dimension-6 (dim-6) operators do not contribute to this violation up to the $\mathcal{O}(\alpha_s)$ accuracy. We demonstrate that the quadratic effects of dim-6 dipole operators, associated with the $Z$ boson, can exert a substantial impact on the breaking of the Lam-Tung relation, surpassing the influence of the quadratic effects of other dim-6 operators and linear effects of dim-8 operators. Furthermore, these effects are prominent not only around the $Z$-boson mass region, but also in the high-invariant mass region of the lepton pairs due to the additional momentum dependence of the dipole interactions, distinguishing them from the nonperturbative effects in QCD~\cite{Brzeminski:2016lwh,Motyka:2016lta,Nefedov:2020ugj,Balitsky:2021fer}. Consequently, the clear deviation from the SM prediction of the Lam-Tung relation around the $Z$-pole may hint for the existence of dim-6 dipole operators from the NP.

%================================================================================================
\vspace{3mm}
\noindent {\bf Angular coefficients.}
%================================================================================================
\label{sec:violation}
The angular distributions of the leptons in the Drell-Yan process can be expanded using harmonic polynomials together with dimensionless angular coefficients $A_i$ in the Collins-Soper (CS) frame~\cite{Collins:1977iv},
\bea
&&\frac{d\sigma}{dp_T^{\ell\ell}dy_{\ell\ell}dm_{\ell\ell\ }^2d\cos{\theta}d\phi\ } \nonumber \\
&=&\frac{3}{16\pi}\frac{d\sigma}{dp_T^{\ell\ell}dy_{\ell\ell}dm_{\ell\ell}^2} \left\{\left(1+\cos^2{\theta}\right)+\frac{1}{2}A_0\left(1-3\cos^2{\theta}\right) \right. \nonumber \\ 
&& +A_1\sin{2\theta}\cos{\phi}+\frac{1}{2}A_2\sin^2{\theta}\cos{2\phi}+A_3\sin{\theta}\cos{\phi} \nonumber \\
&&  +A_4 \cos{\theta}  +A_5\sin^2{\theta}\sin{2\phi}+A_6\sin{2\theta}\sin{\phi}  \nonumber \\
&&  \left. +A_7\sin{\theta}\sin{\phi} \right\} \;,
\label{eq:expansion}
\eea
where $\theta$ and $\phi$ represent the polar and azimuthal angles of the lepton in the CS frame, and $y_{\ell\ell}$ and $m_{\ell\ell\ }$ denotes the rapidity and invariant mass of the lepton pairs, respectively. Due to the orthogonality of the $P_l(\cos\theta,\phi)$ polynomials, the angular coefficients $A_i$ can be obtained by taking the moment of the corresponding polynomial,
\bea
\left\langle P_l\left(\cos{\theta},\phi\right)\right\rangle=\frac{\int P_l\left(\cos{\theta},\phi\right)d\sigma d\cos{\theta}d\phi}{\int d\sigma d\cos{\theta}d\phi} \;.
\label{eq:harmonics}
\eea
Therefore, the angular coefficients $A_0$ and $A_2$ are obtained as follows:
\bea
A_0=4-10\left\langle \cos^2\theta \right\rangle, \quad
A_2=10\left\langle \sin^2\theta\cos2\phi\right\rangle \;. 
\label{eq:A0A2}
\eea

\begin{figure}[htb]
	\begin{center}
		\includegraphics[width=1.\linewidth]{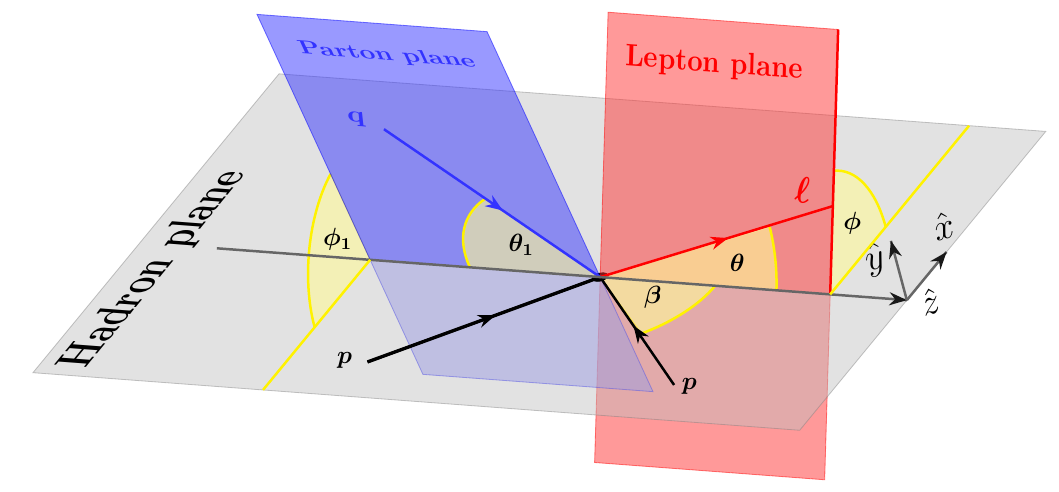}
	\end{center}
	\caption{The Collins-Soper frame, with  planes of parton, leptons and hadrons causing the angular relation of eq.~\eqref{eq:angularrelation}.
	}
	\label{fig:threeplane}
\end{figure}

The lepton angular distribution can also be expanded in the center-of-mass (CM) frame of the lepton pairs, and the differential cross-section for the Drell-Yan process, after including SMEFT effects up to order $1/\Lambda^4$, can have the following form,
\bea
\frac{d\sigma}{d\Omega} = a\cos{\hat{\theta}}+b\cos^2{\hat{\theta}}+c\cos^3{\hat{\theta}}+d \;,
\label{eq:generalM}
\eea
where $\hat{\theta}$ denotes the angle between the negative charged lepton and the incoming quark in the CM frame.  It has been demonstrated that $\hat{\theta}$ in the CM frame can be expressed in terms of the observables $\theta$ and $\phi$ in the CS frame~\cite{Peng:2015spa}. This relationship is given by,
\bea
\cos{\hat{\theta}}=\cos{\theta}\cos{\theta_1}+\sin{\theta}\sin{\theta_1}\cos{\left(\phi-\phi_1\right)},
\label{eq:angularrelation}
\eea
where $\theta_1$ and $\phi_1$ are the polar and azimuthal angles of incoming parton in the CS frame, as illustrated in Fig.~\ref{fig:threeplane}. 
By substituting the above relation into Eq.~\eqref{eq:generalM} and utilizing Eq.~\eqref{eq:A0A2}, we can express the angular coefficients $A_0$ and $A_2$ with the general cross section coefficients $a,b,c,d$ as follows,
\bea
A_0&=&\bigg\langle\frac{2 (d-b)+4b  \sin^2{\theta_1} }{b+3 d} \bigg\rangle,  \nonumber \\
A_2&=&\bigg\langle\frac{4 b \sin^2 { \theta_1} \cos{ 2 \phi_1 }}{b+3 d} \bigg\rangle \;. 
\eea
When the parton plane and the hadron plane coincide, {\it i.e.}, $\phi_1=0$, which is referred to as "coplanarity" and can be maintained up to $\mathcal{O}(\alpha_s)$ accuracy, the $A_0-A_2$ relation becomes simple,
\bea
A_0-A_2= \bigg\langle 2\frac{d-b}{b+3 d} \bigg\rangle \;.
\label{eq:violate}
\eea
Hence, the breaking of Lam-Tung relation can occur for the coplanarity case with $b\neq d$ or the non-coplanarity case ({\it i.e.}, $\phi_1\neq 0$). This implies that the cross section from the SM up to $\mathcal{O}(\alpha_s)$ does not violate this relation, as indicated by the qualitative analysis in Eq.~\eqref{eq:violate}, owing to $b=d$.
Consequently, the leading violation of $A_0=A_2$ in the SM would originate from the non-coplanarity effects of hadron and parton planes, characterized by $\phi_1\neq 0$, which can be induced by the perturbative QCD interaction at  $O(\alpha_s^2)$ and beyond~\cite{Chang:2017kuv}   or by the nonperturbative effects arising from the intrinsic transverse momentum of the incoming parton~\cite{Brzeminski:2016lwh,Motyka:2016lta,Nefedov:2020ugj}. However, even with the inclusion of the most precise theoretical calculations in the $O(\alpha_s^3)$ accuracy~\cite{Gauld:2017tww} and the NLO electroweak correction~\cite{Frederix:2020nyw}, the ATLAS regularized data at the high-$p_T^{\ell\ell}$ region still cannot be described by the SM. To address this discrepancy, in this Letter, we demonstrate that this discrepancy in angular coefficients of the Drell-Yan pairs could be explained by the inclusion  of the electroweak dipole operators of the $Z$-boson at $\mathcal{O}(\alpha_s)$, with $\phi_1=0$.
These physics effects can also be clearly distinguished from the nonperturbative QCD effects, and other potential sources of NP in the Drell-Yan process.

%================================================================================================
\vspace{3mm}
\noindent {\bf Lam-Tung relation breaking in the SMEFT.}
%================================================================================================
The lagrangian that contributes to the neutral Drell-Yan process in SMEFT up to $\mathcal{O}(1/\Lambda^4)$ is given by:
\bea
\mc{L}_\text{SMEFT}=\mc{L}_\text{SM}+\mc{L}_\text{dim-6}+\mc{L}_\text{dim-8}.
\label{eq:lagrangianSMEFT}
\eea
In the notation of Ref.~\cite{Grzadkowski:2010es}, there exist five types of operators at the dim-6 level that are relevant to our study, {\it i.e.},
$\mc{L}_\text{dim-6}=\mc{L}_{X^2\varphi^2}+\mc{L}_{\psi^2X\varphi}+\mc{L}_{\psi^2\varphi^2D}+\mc{L}_{\psi^2\varphi^3}+\mc{L}_{\psi^4}$. The effects of these dim-6 operators on the kinematic distributions in the Drell-Yan process have been extensively studied in the literature~\cite{Alioli:2018ljm,Boughezal:2023nhe,Grossi:2024tou}, and the contributions from dim-8 operators have been included in Refs.~\cite{Alioli:2020kez,Boughezal:2021tih,Boughezal:2022nof,Li:2022rag}. It has been demonstrated that the effects from operators in $\mathcal{L}_{X^2\varphi^2}$, $\mathcal{L}_{\psi^2\varphi^2D}$, and $\mathcal{L}_{\psi^2\varphi^3}$ only lead to an overall shift in the couplings of the SM, and cannot contribute to the violation of the Lam-Tung relation in the coplanarity case, {\it i.e.,} up to $\mathcal{O}(\alpha_s)$ accuracy. However, while these operators could contribute to $A_0-A_2$ at $\mc{O}(\alpha_s^2)$ and beyond, their corrections remain negligible when considering the constraints from electroweak precision measurements at the $Z$-pole, with a typical value of $(A_0-A_2)_{\rm SM}v^2/{\Lambda^2}C_i$ where $C_i \sim \mc{O}(0.01)$ and $v=246~{\rm GeV}$ for $\Lambda=1~{\rm TeV}$ \cite{Ellis:2020unq}. Therefore, we do not explicitly consider them here. 
The four-fermion operators are much more complicated and can be divided into two subsets $O^{(1)}_{\psi^4}$ and $O^{(2)}_{\psi^4}$. We define the $O^{(1)}_{\psi^4}$ as the operators that can interfere with the SM while the $O^{(2)}_{\psi^4}$ cannot. It is evident that $O^{(1)}_{\psi^4}$ takes the form of $\bar{\psi}_i \gamma^\mu \psi_i \bar{\psi}_j \gamma^\mu \psi_j$, which can be interpreted as an operator derived from integrating out a heavy gauge boson. As a result, these four-fermion interactions will resemble those in the SM and will not violate the Lam-Tung relation up to $\mathcal{O}(\alpha_s)$ accuracy either. In comparison to operators in $\mathcal{L}_{X^2\varphi^2}$, $\mathcal{L}_{\psi^2\varphi^2D}$, and $\mathcal{L}_{\psi^2\varphi^3}$, the impact of $O^{(1)}_{\psi^4}$ on $A_0-A_2$ at $\mathcal{O}(\alpha_s^2)$ and higher orders will be additionally suppressed by the factor $\Gamma_Z /M_Z$ due to the angular coefficients being extracted in the invariant mass region near the $Z$-boson mass, where $\Gamma_Z$ represents the $Z$-boson decay width. Although the typical constraints for these Wilson coefficients are approximately $\mathcal{O}(1)$ with $\Lambda=1$ TeV \cite{Boughezal:2023nhe}, the correction to $A_0-A_2$ can still be safely disregarded when considering all effects.

The remaining four-fermion operator $O^{(2)}_{\psi^4}$ and the dipole operators in $\mathcal{L}_{\psi^2 X \varphi}$ may contribute to the Drell-Yan process at $\mathcal{O}(1/\Lambda^4)$. We will demonstrate below that their effects can lead to a violation of $A_0=A_2$ at $\mathcal{O}(\alpha_s)$ accuracy in   high-$p_T^{\ell\ell}$ region, corresponding to the coplanarity case. The dim-6 operators can only violate the Lam-Tung relation at $\mathcal{O}(1/\Lambda^4)$, which are the quadratic effects from the following operators:
\begin{equation}
	\begin{aligned}
	\mc{L}_{\psi^2 X \varphi}=& + \bar{q}_L \sigma^{\mu \nu}\left(C_{uB} B_{\mu \nu}+ C_{uW} \tau^I W_{\mu \nu}^I\right) \frac{\tilde{\varphi}}{\Lambda^2} u_R \\
	& + \bar{q}_L \sigma^{\mu \nu}\left( C_{dB} B_{\mu \nu}+ C_{dW} \tau^I W_{\mu \nu}^I\right) \frac{\varphi}{\Lambda^2} d_R \\
	& + \bar{\ell}_L \sigma^{\mu \nu}\left( C_{eB} B_{\mu \nu}+ C_{eW} \tau^I W_{\mu \nu}^I\right) \frac{\varphi}{\Lambda^2} e_R+\text {h.c. } \\
	\mathcal{L}_{\psi^4}^{(2)}=& \frac{1}{\Lambda^2}\left\{C_{L e d Q} \bar{\ell}_L^i e_R \bar{d}_R q_L^i +C_{L e Q u}^{(1)} \varepsilon^{i j} \bar{\ell}_L^i e_R \bar{q}_L^j u_R \right.  \\ & \left.+C_{L e Q u}^{(3)} \varepsilon^{i j} \bar{\ell}_L^i \sigma^{\mu \nu} e_R \bar{q}_L^j \sigma_{\mu \nu} u_R +\text { h.c. }\right\} \;, 
	\end{aligned}
	\label{eq:dipole1}
\end{equation}
where $q_L(\ell_L)$,  $u_R$, $d_R$ and $e_R$ represent the left-handed quark (lepton) doublet, right-handed quark, and lepton fields, respectively. $B_{\mu\nu}$ and $W_{\mu\nu}^I$ denote the field strength tensor of $U(1)_Y$ and $SU(2)_L$, respectively, and $\varphi$ represents the Higgs doublet. 

It would be straightforward to qualitatively confirm the breaking effects by calculating the lepton angular distribution for the aforementioned dim-6 operators in the CM frame.  It shows that the cross section from the dipole operators is proportional to $\cos^2\hat{\theta}-1$, which arises from the Wigner function $d_{1,0}^1$. The scalar-type four-fermion operators in $\mathcal{L}_{\psi^4}^{(2)}$ could be obtained after integrating out a heavy scalar, corresponding to the s-wave scattering in the Drell-Yan process, and as a result, it does not depend on $\hat{\theta}$. On the other hand, the scattering amplitude from the tensor-type operator in $\mathcal{L}_{\psi^4}^{(2)}$ would depend on the Wigner function $d_{0,0}^1=\cos\hat{\theta}$. Consequently, these dim-6 operators could violate the Lam-Tung relation in high-$p_T^{\ell\ell}$ region at the $\mathcal{O}(\alpha_s)$ accuracy, as estimated in Eq.~\eqref{eq:violate}.

To make a consistent calculation in the SMEFT framework, up to $\mathcal{O}(1/\Lambda^4)$, it is essential to consider the contribution from the linear effects of dim-8 operators to the $A_0-A_2$ relation.
Following a similar analysis as with the dim-6 operators, in line with the discussions outlined in Ref.~\cite{Alioli:2020kez}, it was found that most of the dim-8 operators do not lead to $A_0\neq A_2$ at the $\mathcal{O}(\alpha_s)$ accuracy. Only seven dim-8 operators with the following form can break the Lam-Tung relation~\cite{Li:2020gnx,Murphy:2020rsh},
\begin{align}
	O_{8, lq\partial 3} &= (\bar{\ell}_L\gamma_{\mu} \overleftrightarrow{D}_{\nu}\ell_L) (\bar{q}_L\gamma^{\mu} \overleftrightarrow{D}^{\nu}q_L), \label{eq:op1}
\end{align} 
where $ \overleftrightarrow{D}_\mu= \overrightarrow{D}_\mu- \overleftarrow{D}_\mu$ and $D_\mu$ is a covariant derivative. The remaining operators can be obtained by substituting the $\ell_L$ or $q_L$ with the fermion singlets or by inserting the Pauli matrices $\sigma^I$. The explicit calculation in the CM frame shows that $d\sigma_\text{dim-8}/d\Omega \sim \cos\hat{\theta}(\cos\hat{\theta}\pm 1)^2$. 
It arises from the interference between the Feynman diagram that contains the above operator with its contribution proportional to  $\cos\hat{\theta} \times d_{1,\pm 1}^1$, and the SM diagram, whose contribution is proportional to the Wigner function $d_{1,\pm 1}^1=(1\pm \cos\hat{\theta})/2$. Therefore, the coefficient $d=0$ in Eq.~\eqref{eq:generalM} for the aforementioned dim-8 operators, indicating that the Lam-Tung relation can be broken at $\mathcal{O}(\alpha_s)$, cf. Eq.~\eqref{eq:violate}. 

While numerous dim-6 and dim-8 operators could contribute to the breaking of $A_0=A_2$ in high-$p_T^{\ell\ell}$ region at the $\mathcal{O}(\alpha_s)$ accuracy, the dominant contribution to this violation is expected to come from the dipole operators relevant to the $Z$-boson, given that the angular coefficients are extracted in the invariant mass region near the $Z$-boson mass as discussed before. Consequently, the contributions from the four-fermion dim-6 ($O^{(2)}_{\psi^4}$) and dim-8 operators will be highly suppressed, as compared to the $Z$-boson dipole operators by the ratio $(\Gamma_Z/M_Z)^2$. 
Therefore, in this paper, we focus solely on the dipole operators and disregard the contributions of other operators in the following analysis.

After the electroweak symmetry breaking with $\langle\varphi\rangle=v/\sqrt{2}$, the dipole operators in Eq.~\eqref{eq:dipole1} can be written as
\begin{equation}
	\begin{aligned}
	\mc{L}_{\psi^2 X \varphi} \subset  & \frac{v}{\sqrt{2}\Lambda^2} \left[  \bar{u}_L \sigma^{\mu \nu} u_R \left( C_{uA} A_{\mu \nu} + C_{uZ} Z_{\mu \nu}\right)  \right. \\
	&  + \bar{d}_L \sigma^{\mu \nu} d_R \left( C_{dA} A_{\mu \nu} + C_{dZ} Z_{\mu \nu}\right) \\
	& \left. + \bar{e}_L \sigma^{\mu \nu}  e_R \left( C_{eA} A_{\mu \nu} + C_{eZ} Z_{\mu \nu}\right) +\text {h.c. } \right] \;, 
	\end{aligned}
	\label{eq:dipole2}
\end{equation}
where the coefficients $C_{eA}=  c_W  C_{eB} - s_W C_{eW}  $ and $C_{eZ}= -c_W C_{eB}  - s_W C_{eW}  $, with $c_W (s_W)$ represents the cosine (sine) of the weak mixing angle. Similar definitions also apply for $C_{uA}, C_{dA}, C_{uZ}, C_{dZ}$. By focusing on the invariant mass region around the $Z$-pole with $m_{\ell\ell}\in [80,100]$ GeV, we can neglect the contributions from the photon~\footnote{Photon dipole couplings are also tightly constrained by the anomalous magnetic and electric dipole moments of particles~\cite{Aebischer:2021uvt,ParticleDataGroup:2024cfk}. The mixing of photon and $Z$ dipole operators at the loop level can further tighten constraints on $Z$ dipole interactions, depending on the theoretical assumptions made in the analysis~\cite{Aebischer:2021uvt}.}. Therefore, the cross section from the NP is solely a function of $\{C_{eZ}, C_{uZ},C_{dZ} \}$.

%================================================================================================
\vspace{3mm}
\noindent {\bf Numerical results and discussion.}
%================================================================================================
We now present the projected constraints on the dipole operators from the $A_0-A_2$ measurements in the Drell-Yan process at the LHC, in accordance with Eq.~\eqref{eq:A0A2}. To calculate the breaking effects, we adopt the $G_\mu$-scheme with the following electroweak parameters~\cite{Workman:2022ynf}: $M_Z^{\rm os}=91.1876\ \text{GeV}$, $\Gamma_Z^{\rm os}=2.4952\ \text{GeV}, M_W^{\rm os}= 80.385\ \text{GeV},\Gamma_W^{\rm os}= 2.085 \ \text{GeV}$ 
and $G_\mu = 1.16638 \cdot 10^{-5}\ \text{GeV}^{-2}$, along with the {\ttfamily CT18NNLO} PDF set\cite{Hou:2019efy}.

\begin{figure}[htb]
	\begin{center}
		\includegraphics[width=0.85\linewidth]{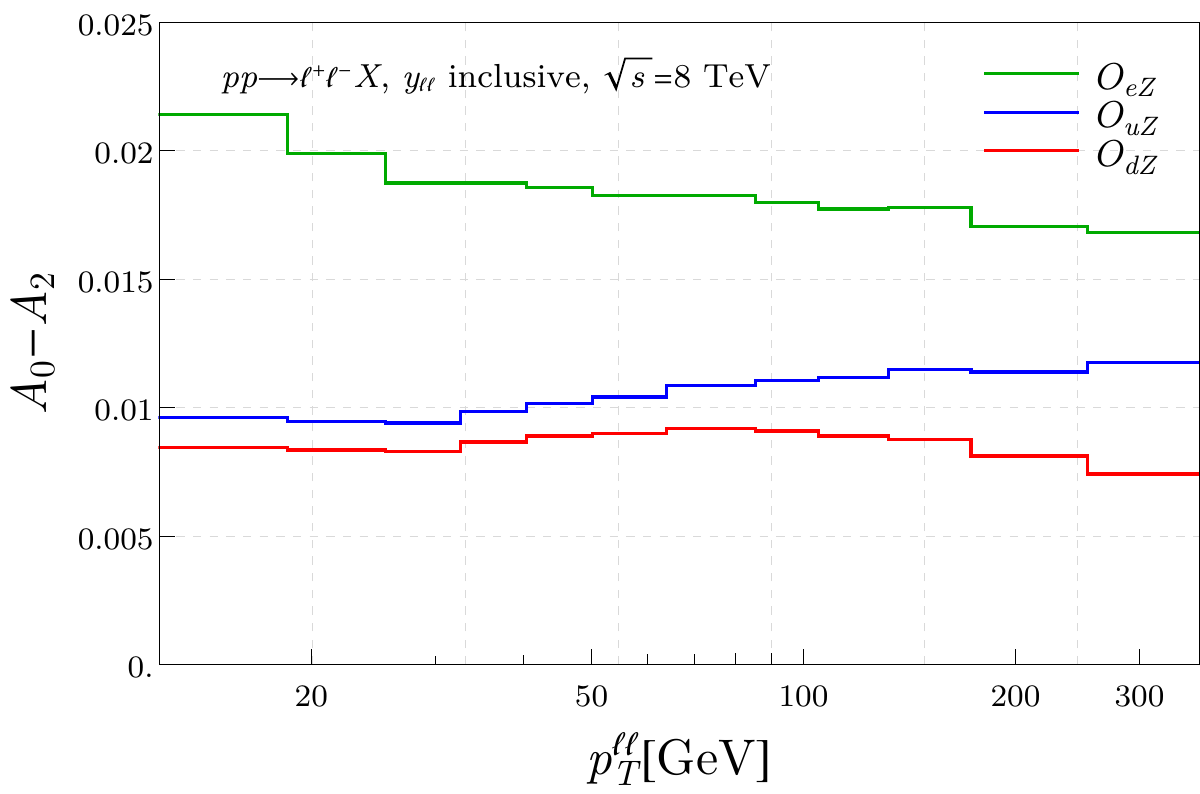}
	\end{center}
	\caption{The contribution of dipole operators to the difference of angular coefficients $A_0-A_2$ as a function of $p_T^{\ell\ell}$ in the region $m_{\ell\ell}\in [80,100]$ GeV at $\sqrt{s}=8~{\rm TeV}$ LHC, with $C_i=1$ and $\Lambda=1~{\rm TeV}$.}
	\label{fig:A0A2-low-s}
\end{figure}

Figure~\ref{fig:A0A2-low-s} presents the predicted $A_0-A_2$ from the dipole operators relevant to the $Z$ boson with $C_i=1$ and $\Lambda=1~{\rm TeV}$, at $\sqrt{s}=8~{\rm TeV}$ with the dilepton invariant mass $m_{\ell\ell}\in [80,100]$ GeV. The $\mathcal{O}(\alpha_s^3)$ prediction in the SM~\cite{Gauld:2017tww} is used as the normalization for the angular coefficients in Eq.~\eqref{eq:expansion}. The plot clearly indicates that the breaking effects induced by those dipole operators could be significant, suggesting that they could be a strong signal of the $Z$-boson dipole interaction among potential sources of NP in the Drell-Yan process.

Next, we conduct a $\chi^2$ analysis to constrain the Wilson coefficients of dipole operators from the $\Delta\equiv A_0-A_2$ measurements at the LHC,
\bea
\chi^2=\sum_i \frac{\left(\Delta _{\exp }-\Delta _{\text{SM}}-\Delta _{\text{SMEFT}}\right){}^2}{\delta \Delta _{\exp }^2+\delta \Delta _{\text{Theo}}^2} \;,
\eea
where we sum over the $p_T^{\ell\ell}$ bins. The $\Delta _{\exp }$, $\Delta _{\text{SM}}$, and $\Delta _{\text{SMEFT}}$ represent, respectively, the experimental measurements at the LHC, the SM prediction at $\mathcal{O}(\alpha_s^3)$ accuracy, and the contributions from dipole operators at $\mathcal{O}(\alpha_s)$ accuracy for the $i$-th $p_T^{\ell\ell}$ bin. The $\delta \Delta _{\exp }$ encompasses the total uncertainty from experimental measurements, including statistical and systematic errors, while the  $\delta\Delta _{\text{Theo}}$ corresponds to the scale uncertainty of the QCD calculation. In this work, we have taken the transverse mass of $Z$-boson as the canonical scale.

\begin{figure}[htb]
	\begin{center}
		\includegraphics[width=0.8\linewidth]{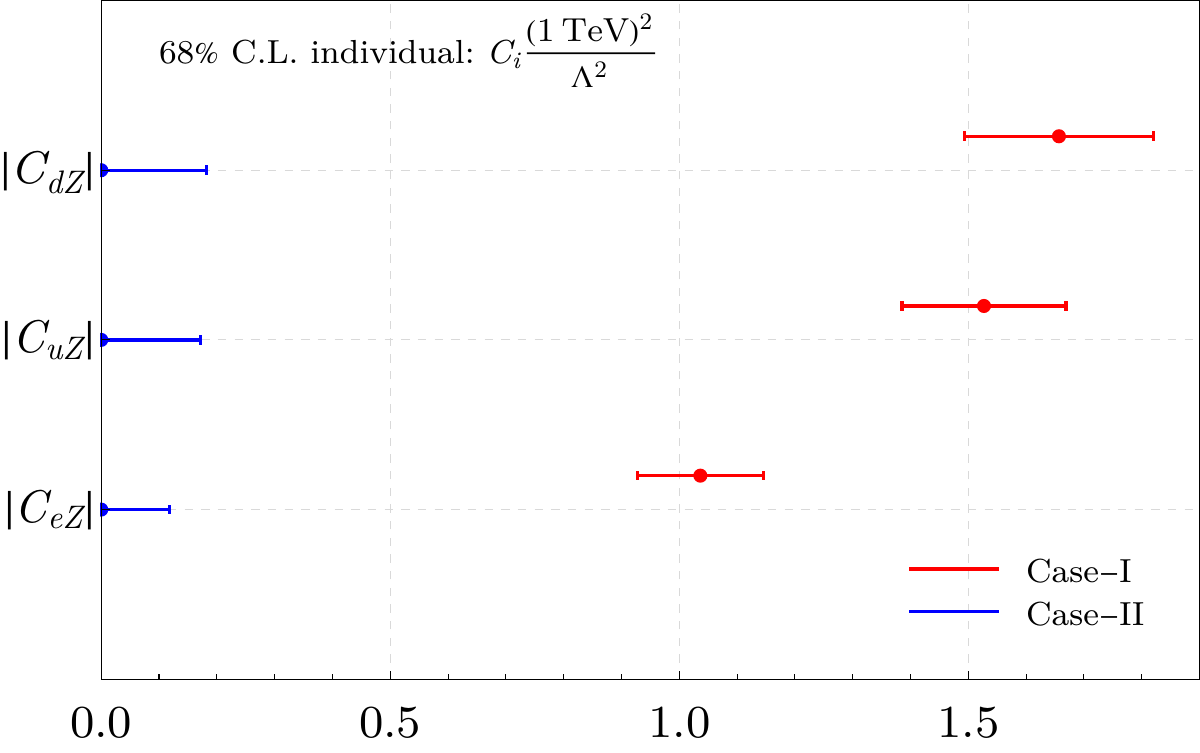}
	\end{center}
	\caption{The individual bounds for the Wilson coefficients of dim-6 dipole operators at 68\% C.L. with $\Lambda=1~{\rm TeV}$.
	}
	\label{fig:individual}
\end{figure}

To derive the individual bounds for the dim-6 dipole operators given in Eq.~\eqref{eq:dipole2}, we consider two scenarios in this Letter:
\begin{enumerate} [I.]
	\item $\Delta _{\exp }$ corresponds to the current ATLAS measurements with regularized data, at $\sqrt{s}=8~{\rm TeV}$~\cite{ATLAS:2016rnf}, incorporating both the theoretical and experimental uncertainties in $\chi^2$ analysis~\cite{ATLAS:2016rnf};
	\item Pseudo-experiments at the $\sqrt{s}=14~{\rm TeV}$ LHC with $3000~{\rm fb}^{-1}$ (HL-LHC) are conducted with the central values consistent with $\mathcal{O}(\alpha_s^3)$ prediction in the SM. Here, we only consider the statistical error, with rescaling from Ref.~\cite{ATLAS:2016rnf}.
\end{enumerate}

\begin{figure}[htb]
	\begin{center}
		\includegraphics[width=0.9\linewidth]{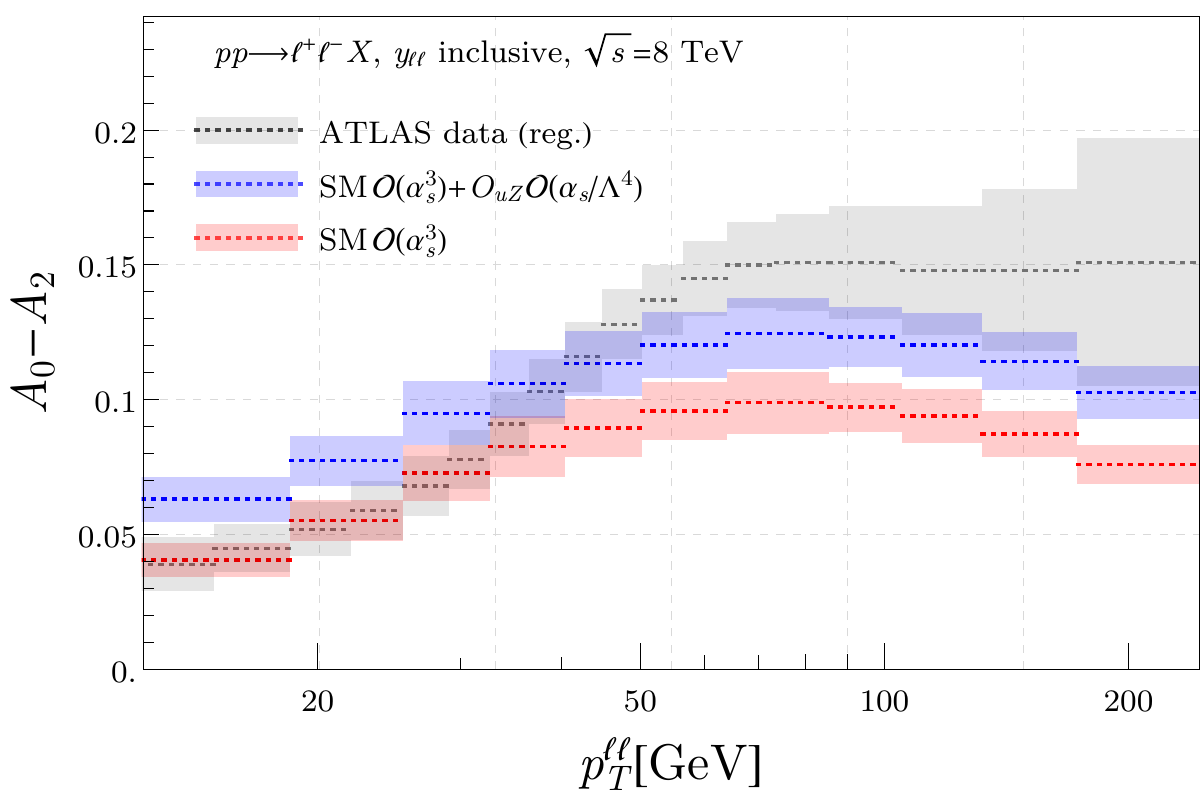}
	\end{center}
	\caption{The distribution of angular coefficient $A_0-A_2$ as a function of $p_T^{\ell\ell}$ at $\sqrt{s}=8~{\rm TeV}$ LHC. The ATLAS regularized data~\cite{ATLAS:2016rnf} (gray band) are compared to the $\mathcal{O}(\alpha_s^3)$ prediction in the SM~\cite{Gauld:2017tww} (red band) and the combined prediction  from the SM and the operator $\mathcal{O}_{uZ}$ with the fitted Wilson coefficient in Fig.~\ref{fig:individual} (blue band).
	}
	\label{fig:current}
\end{figure}

For case-I, the red lines in Fig.~\ref{fig:individual} show the allowed range of the Wilson coefficients of those dipole operators, when considering them one at a time, at the 68\% confidence level (C.L.) 
To estimate the goodness of the $\chi^2$ analysis, we calculate the $\chi^2_{\rm min}/{\rm dof}$ for those three operators one at a time, which are found to be 1.98 for $\mathcal{O}_{eZ}$, 1.61 for  $\mathcal{O}_{uZ}$, and 1.73 for  $\mathcal{O}_{dZ}$. This suggests that the ATLAS regularized data can be better described by a non-vanishing contribution from the operator $\mathcal{O}_{uZ}$, which can be easily seen by comparing the shape of data and those curves presented in Fig.~\ref{fig:A0A2-low-s}. In Fig.~\ref{fig:current}, we compare $A_0-A_2$ as a function of $p_T^{\ell\ell}$ from the ATLAS regularized data (gray band), the SM prediction at $\mathcal{O}(\alpha_s^3)$ accuracy (red band), and the combined prediction from the SM and the dipole operator $\mathcal{O}_{uZ}$ with the fitted $C_{uZ}$ value (blue band), at $\sqrt{s}=8~{\rm TeV}$. It is evident that this discrepancy in the $A_0-A_2$ measurement could be well explained after including the contribution of $\mathcal{O}_{uZ}$. However, we cannot conclusively attribute the potential discrepancy in the $A_0-A_2$ measurement to the contribution of the operator $O_{uZ}$, given the substantial experimental uncertainties. 

For case-II, we calculate the cross section of the SM presented in Eq.~\eqref{eq:expansion} at $\sqrt{s}=14~{\rm TeV}$ with $\mathcal{O}(\alpha_s^2)$ accuracy using MadGraph5\_aMC@NLO~\cite{Alwall:2014hca} and use the ratio $\kappa=\sigma(\mathcal{O}(\alpha_s^3))/\sigma(\mathcal{O}(\alpha_s^2))$  presented in Ref.~\cite{Gauld:2017tww} for $\sqrt{s}=8~{\rm TeV}$ to estimate the $\mathcal{O}(\alpha_s^3)$ correction at $\sqrt{s}=14~{\rm TeV}$, as this ratio demonstrates insensitivity to the collider energy across the range from 8 TeV to 13 TeV \cite{Boughezal:2015ded}. Updating these results with the exact $\mathcal{O}(\alpha_s^3)$ prediction, instead of making the above approximation, of the SM contribution at $\sqrt{s}=14~{\rm TeV}$ would be straightforward. However, it is beyond the scope of this paper.

It shows that even if the experimental measurements are consistent with the SM prediction in the future, the measurements of the breaking effects of Lam-Tung relation would still provide crucial information on the dipole operators, without assuming other NP effects in the Drell-Yan process, as indicated by the blue lines in Fig.~\ref{fig:individual}. These results are comparable to the limits from the $m_{\ell\ell}$ distribution of the Drell-Yan process~\cite{Boughezal:2021tih} and the Higgs and electroweak precision measurements~\cite{Escribano:1993xr,Alioli:2018ljm,daSilvaAlmeida:2019cbr,Bonnefoy:2024gca} when one operator is considered at a time. However, the conclusions drawn from these measurements heavily depend on the theoretical assumptions made in the analysis, which may not directly yield robust constraints for these NP effects. In contrast, the measurements of $A_0-A_2$ proposed in this article do not depend on other dim-6 and dim-8 SMEFT operators, except for the dipole operators listed in Eq.~\eqref{eq:dipole2}. 

\begin{figure}[htb]
	\begin{center}
		\includegraphics[width=0.9\linewidth]{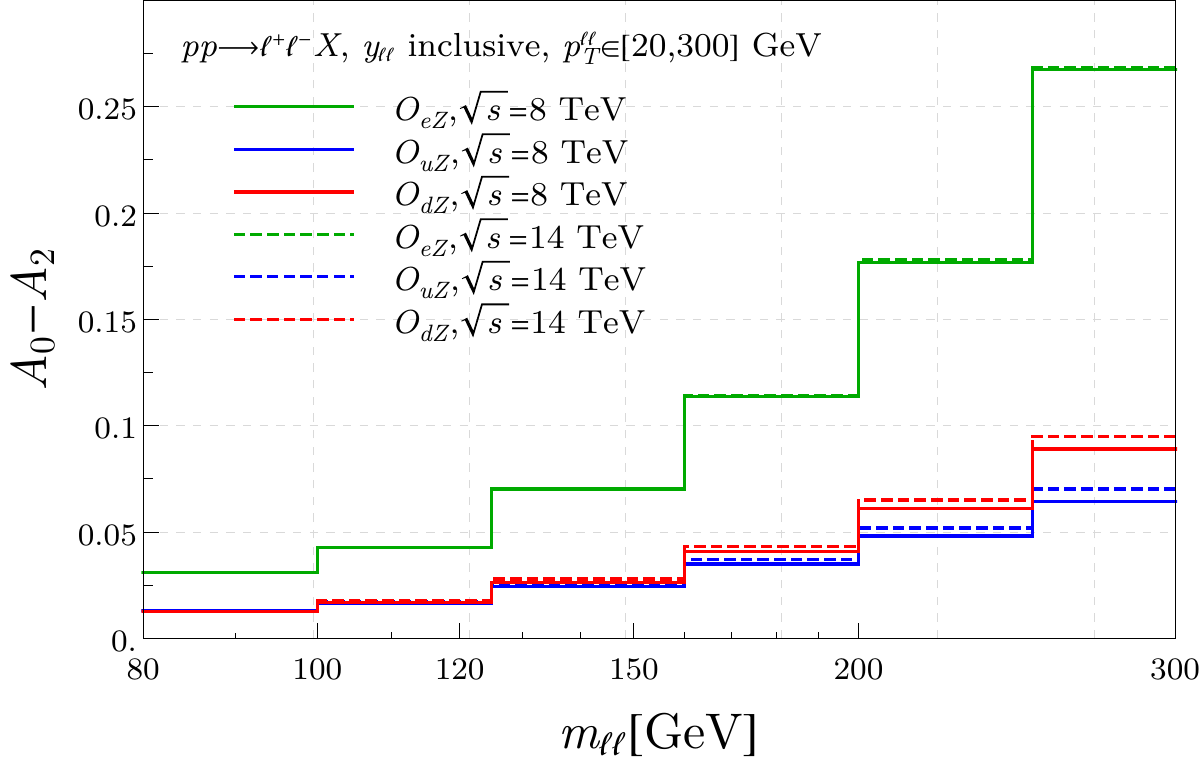}
	\end{center}
	\caption{The distribution of angular coefficient $A_0-A_2$ as a function of $m_{\ell\ell}$ at $\sqrt{s}=8~{\rm TeV}$ and $\sqrt{s}=14~{\rm TeV}$ LHC, with $C_i=1$ and $\Lambda=1~{\rm TeV}$.
	}
	\label{fig:mass-dis}
\end{figure}

Finally, it is important to note that these NP effects for the $A_0-A_2$ measurements would be prominent not only when the Drell-Yan pair invariant mass is around the $Z$-boson mass, but also become more significant in the high-invariant mass region of lepton pairs due to the additional momentum dependence of the dipole interactions, see Fig.~\ref{fig:mass-dis}. We have taken into account the contributions from both the photon and $Z$-boson at $\mathcal{O}(\alpha_s^3)$ using the rescaling method in the normalization of the angular coefficients in the off $Z$-pole region. Therefore, to verify these effects and distinguish them from nonperturbative QCD effects, we strongly recommend to also measure this relation far away from  the $Z$-boson mass window. If the new measurements are inconsistent with the SM prediction, the discrepancy in the Lam-Tung relation around the $Z$-pole would be a significant signal of the electroweak dipole interactions, induced by some NP interactions. Furthermore, these NP effects may also contribute to the fermion $g-2$ indirectly and induce sizable transverse spin asymmetry at colliders. Therefore, their effects could be further crosschecked by low-energy experiments~\cite{Aebischer:2021uvt}, and spin observable at future lepton colliders~\cite{Wen:2023xxc}, Electron-Ion colliders~\cite{Boughezal:2023ooo,Wang:2024zns,Wen:2024cfu} and heavy-ion collider~\cite{Shao:2023bga}.

%================================================================================================
\vspace{3mm}
\noindent {\bf Conclusions.}
%================================================================================================
In this Letter, we present a model-independent investigation of the observed discrepancy in the Lam-Tung relation measurements in the Drell-Yan process at the LHC. We demonstrate that the leading contribution from the SMEFT to the breaking effects in high-$p_T^{\ell\ell}$ region arises from $\mathcal{O}(1/\Lambda^4)$ at the $\mathcal{O}(\alpha_s)$ accuracy in QCD interaction and is dominated by the dim-6 dipole interactions relevant to the $Z$ boson in the $Z$-boson mass window. Consequently, the discrepancy in Lam-Tung relation measurements could be a possible signal of the electroweak dipole interactions induced by NP, and this conclusion is independent of other potential NP effects in the Drell-Yan process. Furthermore, the NP effects induced by these momentum dependent dim-6 dipole operators would become more significant when $m_{\ell\ell}$ is much larger than $M_Z$, allowing them to be clearly distinguished from the nonperturbative QCD effects, which could also potentially contribute to the breaking of Lam-Tung relation. Even if the discrepancies were to vanish in future measurements, the $A_0$–$A_2$ measurements would remain a powerful tool for constraining $Z$-boson dipole interactions.

\vspace{3mm}
\noindent{\bf Acknowledgments.}
X. L. is supported by the National Natural Science Foundation of China under grant No. 11835013. B. Y. 
is supported in part by the National Science Foundation of China under Grants No.~12422506, the IHEP under grant No.~E25153U1 and CAS under grant No.~E429A6M1. C.-P. Y.
is supported by the U.S. National Science Foundation
under Grant No. PHY-2310291.

\bibliographystyle{apsrev}
\bibliography{refs}

\end{document}